\documentclass[aps,prb,twocolumn,amsmath,amssymb,showpacs]{revtex4-1}
\usepackage{color}
\usepackage{graphicx}

\begin{document}
\title{Weak localization of Dirac fermions in graphene \\ beyond the diffusion regime}

\author{M.\,O.\,Nestoklon}
\author{N.\,S.\,Averkiev}
\author{S.\,A.\,Tarasenko}
\affiliation{A.F. Ioffe Physical-Technical Institute of the RAS, 194021 St. Petersburg, Russia}

\begin{abstract}We develop a microscopic theory of the weak localization of 
  two-dimensional massless Dirac fermions which is valid in the whole range of 
  classically weak magnetic fields. The theory is applied to calculate magnetoresistance 
  caused by the weak localization in graphene and conducting surfaces of bulk 
  topological insulators.
\end{abstract}


\maketitle

\section{Introduction}
The two-dimensional layer of graphene is a most interesting object for the theoretical 
and experimental study of quantum phenomena. It has quite high mobility tunable by a 
gate voltage and a unique band structure similar to the energy spectrum of massless 
Dirac fermions.\cite{Geim07,Peres} Such a quasi-relativistic nature of free carriers 
affects the transport phenomena including the weak localization. The weak localization 
(antilocalization) is caused by the constructive (destructive) interference of electron 
waves traveling in opposite directions.\cite{Hikami} The interference is suppressed by 
an external magnetic field, which leads to a magnetoresistance in classically weak 
fields. The anomalous magnetoresistance in graphene is the subject of intensive 
experimental~\cite{Morozov,Wu,Gorbachev,Tikhonenko08,Ki,Yan,Tikhonenko09,Moser} and 
theoretical~\cite{McCann,Morpurgo,Ostrovsky,Kechedzhi,Robinson,Oppen,Imura} study during 
the last few years. It was experimentally shown that increasing the carrier density and 
decreasing the temperature leads to a transition from weak antilocalization to weak 
localization in graphene.\cite{Tikhonenko09} Such a behavior is attributed to the 
trigonal warping of electron spectra in valleys, which suppresses antilocalization, 
and intervalley scattering, which restores localization.\cite{McCann} 

Depending on the ratio between the magnetic length $\ell_B$ and the mean free path $\ell$
one distinguishes between two regimes of weak localization. In very low magnetic fields 
($\ell_B\gg\ell$), the main contribution to the magnetoresistance comes from large 
diffusion-like trajectories of electrons with the size $L\gg\ell$. This is the diffusion 
regime of weak localization. With the field increase ($\ell_B \sim \ell$), the role of 
large trajectories is suppressed and the weak-localization correction to conductivity is 
determined by electron trajectories with few scatterers. Such a non-diffusion regime is 
realized in high-mobility structures at rather small magnetic fields.\cite{Kavabata,Zuzin,Dyakonov} 
To correctly extract kinetic parameters of carriers, such as the phase breaking time, 
from the magnetoresistance measurements one has to analyze experimental data in the whole
range of classically weak magnetic fields. Previous calculations of magnetoresistance in 
graphene were carried out only for the diffusion regime though experimental data in 
Ref.~\cite{Tikhonenko09} indicate that the magnetic length may reach the mean free path 
in the magnetic field as small as $30\,$mT. The goal of this work is to develop the 
theory of weak localization beyond the diffusion regime and derive the quantum 
correction to conductivity.

The developed theory can be also applied to two-dimensional systems formed at surfaces 
of bulk topological insulators, such as Bi$_2$Se$_3$, Bi$_2$Te$_3$, and 
Bi$_2$Te$_2$Se (see Refs.~[\onlinecite{Moore10},\onlinecite{Ren10}] and references therein), 
and in HgTe quantum wells of critical width.\cite{Tkachov_nature}
The excitations in these systems are described by the effective Hamiltonian similar 
to one for the free carriers in graphene. We note 
that the Dirac cones in graphene are situated at the $K$ points of the two-dimensional 
Brillouin zone and degenerate in spin. In contrast, the similar energy spectrum of
carriers in topological insulators is formed by spin-orbit interaction, with the 
Dirac point being situated at the $\Gamma$ point of the two-dimensional Brillouin 
zone. Experimental data indicate that the magnetoresistance in Bi$_2$Te$_3$ 
(Ref.~\onlinecite{He}) and Bi$_2$Si$_3$ (Ref.~\onlinecite{Chen}) 
as well as in HgTe quantum wells\cite{Olshanetsky} is positive (antilocalization) as 
expected for massless Dirac fermions.

\section{Origin of weak antilocalization in graphene.}
Technically, the calculation of quantum corrections to conductivity involves the 
computation of Cooperons.\cite{Hikami} In the case of multi-valley systems with 
negligible spin-orbit coupling, such as graphene or silicon, the Coorepons are derived 
from an integral equation with the kernel 
$\left\langle V_{\alpha\beta}({\bf k}, {\bf k}') V_{\gamma\delta}(-{\bf k}, -{\bf k}')\right\rangle$, 
where the indices $\alpha$, $\beta$, $\gamma$, and $\delta$ enumerate valleys, ${\bf k}$
and ${\bf k}'$ are the electron wave vectors measured from the valleys centers, and the 
angle brackets denote averaging over the positions of scatterers. We note that in a 
single-valley system due to the time inversion symmetry, one may take the orbital wave 
functions of free electrons in the form $\psi_{{\bf k}}=\psi_{-{\bf k}}^*$, which yields 
$V(-{\bf k}, -{\bf k}')=V({\bf k}', {\bf k})$. Then, the correlator 
$\left\langle V({\bf k}, {\bf k}') V(-{\bf k}, -{\bf k}')\right\rangle$ equals to the 
correlator $\left\langle \left| V({\bf k}, {\bf k}')  \right|^2 \right\rangle$ determine 
the single-particle transport. It leads to a diffusion pole in the Cooperon equation and,
therefore, to an enhancement of backscattering (weak localization). In contrast, 
$V_{\alpha\alpha}(-{\bf k}, -{\bf k}')\neq V_{\alpha\alpha}({\bf k}', {\bf k})$ generally
in multi-valley systems, there are no diffusion poles in intra-valley Cooperons, and weak
localization is absent in the approximation of independent valleys.\cite{comment} 
However, the special form of scattering amplitude may result in a diffusion pole even in 
the one-valley model. 

To consider this phenomenon for graphene we neglect the small trigonal warping of 
electron spectrum and take the effective Hamiltonian describing electron states in each 
valley in the form
\begin{equation}\label{Hamiltonian}
\hat{H} = v \left( \begin{array}{cc}
0   & p_x - i p_y   \\
p_x + i p_y & 0  
\end{array}\right) \:.
\end{equation}
Here, $v$ is electron velocity and ${\bf p}=(p_x,p_y)$ is the momentum operator.
The wave functions of conduction band may be chosen in the form
\begin{equation}\label{psi}
\Psi_{{\bf k}}({\bf r}) = 
 \frac1{\sqrt2}\left(\begin{array}{c}
   1  \\ {\rm e}^{i\varphi} 
\end{array}\right) {\rm e}^{i {\bf k} \cdot {\bf r} } \:,
\end{equation}
where $\varphi$ is the polar angle of the in-plane wave vector ${\bf k}$. We consider 
$n$-doped graphene and assume that the valence-band states do not contribute to 
low-temperature conductivity.

The intra-valley electron scattering from a short-range impurity ($A_1$ symmetry) is 
described by the Hamiltonian
\begin{equation}\label{H_imp}
\delta\hat{H} = a \left(\begin{array}{cc} 1 & 0 \\ 0 & 1  \end{array}\right) \delta({\bf r} - {\bf r}_j) \:,
\end{equation}
where ${\bf r}_j$ is the impurity position. It follows from Eqs.~(\ref{psi}) 
and~(\ref{H_imp}) that the matrix element of scattering has the form 
\begin{eqnarray}\label{scattering}
V( {\bf k}', {\bf k}) &=&  a \frac{1+ {\rm e}^{i(\varphi-\varphi')}}{2} {\rm e}^{i({\bf k}-{\bf k}')\cdot {\bf r}_j}    \\
&\propto&  {\rm e}^{i (\varphi-\varphi')/2} \cos [(\varphi-\varphi')/2]  \nonumber \:.
\end{eqnarray}
One can see that the direct back scattering from an impurity is suppressed and, what is 
more important for quantum effects, the scattering introduces the phase 
$(\varphi-\varphi')/2$ to the electron wave function. Therefore, an electron traveling 
clockwise along a closed path and finally scattered back gains the additional phase 
$\pi/2$ while the electron traveling in the opposite direction gains the phase 
$-\pi/2$. The phase shift of $\pi$ between these two waves results in a destructive 
interference and, hence, in the antilocalization of carriers. 

We note that other forms of scattering amplitude, e.g.,
$V({\bf k}',{\bf k}) \propto a + b \exp[i(\varphi-\varphi')]$ with $a \neq b$, lead to 
the phase gain which depends on the particular trajectory even for closed paths. 
Averaging over the trajectories destroys the wave interference and results in no quantum
corrections to conductivity (see also Refs.~\onlinecite{Morozov,McCann,Morpurgo}).  
Similar arguments apply for a surfaces of bulk topological insulators
and HgTe QWs with the Dirac-like spectrum of carriers.

\section{Calculation of magnetoconductivity.} 
To calculate the interference corrections to conductivity in the perpendicular magnetic 
field we use the diagram technique. The retarded and advanced Green functions of 
conduction electrons in one valley in graphene have the form
\begin{equation}\label{Green}
G^{R,A}({\bf{r}},{\bf{r}}') = \sum_{N, k_y} \frac{ \Psi_{N, k_y}({\bf r}) \Psi_{N, k_y}^\dag({\bf r}')}{\varepsilon_F - \varepsilon_N \pm i \hbar/(2\tau) \pm i \hbar/(2\tau_{\phi}) } \:.
\end{equation}
Here, $\varepsilon_F = v \hbar k_F$ is the Fermi energy, $k_F$ is the Fermi wave vector, 
$\varepsilon_N = \hbar\omega_c \sqrt{N}$, $\omega_c = \sqrt{2}v/\ell_B$, $\tau$ is the 
quantum relaxation time, $1/\tau=n_i a^2 k_F/(2\hbar^2 v)$, $n_i$ is the impurity 
density, $\tau_{\phi}$ is the phase breaking time, we assume that $\tau_{\phi} \gg \tau$,
$\Psi_{N, k_y}({\bf r})$ are the two-component wave functions given in the Landau gauge 
by
\[
\Psi_{N, k_y}({\bf r}) = \frac{1}{\sqrt{2}} \left( 
\begin{array}{c}
\psi_{N-1, k_y}({\bf r}) \\ \psi_{N, k_y}({\bf r})
\end{array}
\right) \:,
\]
$\psi_{N, k_y}({\bf r})$ are the standard functions of a two-dimensional charge particle 
in magnetic field, and $N$ and $k_y$ are the quantum numbers. Assuming that the magnetic 
field is classically weak, i.e., $\omega_c \tau \ll 1$, we obtain
\begin{equation}
G^{R,A}({\bf{r}},{\bf{r}}') = \exp{\left[ -i\frac{(x+x')(y-y')}{2\ell_B^2} \right]} G^{R,A}_0( {\bf{r}}-{\bf{r}}' ) \:,
\end{equation}
where $G^{R,A}_0( {\bf{r}}-{\bf{r}}' )$ are the Green functions at zero field,
\[
G^{R,A}_0({\boldsymbol{\rho}}) = 
- \frac{ \exp[-\rho/(2\ell') \pm i ( k_F \rho + \pi/4) ]}{\sqrt{ 2\pi \rho / k_F} \: \hbar v}  \, g^{R,A}({\boldsymbol{\rho}}) \:,
\]
\[
g^{R,A}({\boldsymbol{\rho}}) =  \frac{1 \pm [\boldsymbol{\sigma} \times {\bf n}]_z}{2} \:,
\]
$\ell'=\ell/\left(1+\tau/\tau_{\phi}\right)$, $\ell=v \tau$ is the mean free path, 
$\boldsymbol{\sigma}$ is the vector of Pauli matrices, ${\bf n}=\boldsymbol{\rho}/\rho$ 
is the unit vector pointing along $\boldsymbol{\rho}$, and it is assumed that 
$k_F \rho \gg 1$. We note that the Green functions are matrices $2 \times 2$ owing to the 
matrix form of the Hamiltonian~(\ref{Hamiltonian}). 

The Cooperon ${\cal C}({\bf{r}},{\bf{r'}})$ may be presented as a matrix $4 \times 4$ in 
the basis of direct product of the states $(1,0)^{{\rm T}}$ and $(0,1)^{{\rm T}}$ inside 
the valley. It is found from the integral matrix equation
\begin{equation}\label{Cooperon_eq}
{\cal C}({\bf{r}},{\bf{r'}}) =  w P({\bf{r}},{\bf{r}}') +
\int P({\bf{r}},{\bf{r}}_1) \, {\cal C}({\bf{r}}_1,{\bf{r'}}) \, d{\bf{r}}_1 \:,
\end{equation}
with the kernel
\begin{multline}\label{Cooperon_kernel}
P({\bf{r}},{\bf{r}}') =  w
G^A({\bf{r}},{\bf{r}}') \otimes G^R({\bf{r}},{\bf{r}}') = \\
\frac{P_0({\bf{r}},{\bf{r}}')}2
\left( \begin{array}{cccc}
   1     & in_-  & -in_-   &   n_-^2 \\
 -in_+   &  1    &  -1     & -in_-   \\
  in_+   & -1    &   1     &  in_-   \\
   n_+^2 & in_+  & -in_+   &   1  
\end{array} \right) \:,
\end{multline}
where $w = n_i a^2$ and
\[
P_0({\bf{r}},{\bf{r}}')=\frac{{\rm e}^{-|{\bf r}-{\bf r}'|/\ell'}}{2\pi \ell |{\bf r}-{\bf r}'|} \exp{\left[ -i\frac{(x+x')(y-y')}{\ell_B^2} \right]} \:.
\]
The matrix elements $P_{\nu \mu}({\bf{r}},{\bf{r}}')$ in Eq.~(\ref{Cooperon_kernel}) are 
obtained from the components 
$G^A_{\alpha\beta}({\bf{r}},{\bf{r}}') G^R_{\gamma \delta}({\bf{r}},{\bf{r}}')$ where the 
indices $\nu$ and $\mu$ enumerate the states $(\alpha,\gamma)$ and $(\beta,\delta)$, 
respectively.

A standard method to solve Cooperon equations is to expand the kernel in series of the 
wave functions $\phi_{N k_y}({\bf r})$ of a particle with the charge $2e$ in the 
magnetic field.\cite{Kavabata,Zuzin} To solve matrix Eq.~(\ref{Cooperon_eq}) we modify 
the approach and introduce the basis matrices
\begin{subequations}\label{basisN}
\begin{align}
\Phi_{N \geq 1, k_y}({\bf{r}})=&
\left( \begin{array}{cccc}
   0        &  0        & \phi_{N-1} & 0 \\
   \dfrac{\phi_{N}}{\sqrt2} &  \dfrac{\phi_{N}}{\sqrt2} & 0      & 0 \\
   \dfrac{\phi_{N}}{\sqrt2} & -\dfrac{\phi_{N}}{\sqrt2} & 0      & 0 \\
   0        &  0        &  0 &  \phi_{N+1} 
\end{array} \right) ,
\\
\Phi_{0, k_y}({\bf{r}})=&
\left( \begin{array}{cccc}
   0        &  0        & 0  & 0 \\
   \dfrac{\phi_{0}}{\sqrt2} &  \dfrac{\phi_{0}}{\sqrt2} & 0      & 0 \\
   \dfrac{\phi_{0}}{\sqrt2} & -\dfrac{\phi_{0}}{\sqrt2} & 0      & 0 \\
   0        &  0        & \phi_{0} & \phi_{1} 
\end{array} \right) .
\end{align}
\end{subequations}
One can see that $\int \Phi_{N, k_y}^\dag({\bf{r}}) \Phi_{N', k'_y}({\bf{r}}) d {\bf{r}} 
= \delta_{N,N'} \delta_{k_y,k'_y} \mathbb{I}$ with $\mathbb{I}$ being the unit matrix 
$4 \times 4$. 

Direct calculation shows that the matrix kernel $P({\bf{r}},{\bf{r}}')$ is expanded in 
series of $\Phi_{N, k_y}({\bf{r}})$ as follows
\begin{equation}\label{P_in_basis}
P( {\bf{r}},{\bf{r}}' ) = 
\sum_{N,k_y} \Phi_{N,k_y}({\bf{r}}) P_N \Phi_{N,k_y}^{\dag}({\bf{r}}') \:,
\end{equation}
where
\begin{subequations}\label{Pnn}
\begin{align}
P_{N \geq 1} = & \left(\begin{array}{cccc}
0 & 0 & 0 & 0 \\
0 &  P_{N}^{(0)}                   &  -i\dfrac{P_{N}^{(1)}}{\sqrt2}   & i\dfrac{P_{N+1}^{(1)}}{\sqrt2}  \\
0 & -i\dfrac{P_{N}^{(1)}}{\sqrt2}   &  \dfrac{P_{N-1}^{(0)}}2          &  
\dfrac{P_{N+1}^{(2)}}2  \\
0 & i\dfrac{P_{N+1}^{(1)}}{\sqrt2}  &  \dfrac{P_{N+1}^{(2)}}2          &  \dfrac{P_{N+1}^{(0)}}2 
\end{array}\right) ,
\\
P_{0} = &
\left(\begin{array}{cccc}
0 & 0 & 0 & 0 \\
0 & P_0^{(0)}  &  0  & i \dfrac{P_1^{(1)}}{\sqrt2}  \\
0 & 0      &  \dfrac{P_0^{(0)}}2        & 0 \\
0 & i\dfrac{P_1^{(1)}}{\sqrt2}  &  0             & \dfrac{P_1^{(0)}}2
\end{array}\right) ,
\end{align}
\end{subequations}
$P_N^{(M)}$ are coefficients given by the integrals
\begin{multline}\label{PNM}
P_N^{(M)} = \frac{\ell_B}{\ell} \sqrt{\frac{(N-M)!}{N!}} \\
\times 
\int_0^{\infty} \exp{\left[ - x \frac{\ell_B}{\ell'} - \frac{x^2}{2} \right]}
L_{N-M}^{(M)}(x^2) \, x^M dx \:,
\end{multline}
and $L_{N-M}^{(M)}$ are the Laguerre polynomials. 
Having decomposed the kernel $P( {\bf{r}},{\bf{r}}' )$, one can readily find that the 
solution of Eq.~(\ref{Cooperon_eq}) for the Cooperon has the form
\begin{equation}
{\cal C}({\bf{r}},{\bf{r}}')= \sum_{N,k_y} \Phi_{N,k_y}({\bf{r}}) \left( \mathbb{I} - P_N \right)^{-1} P_N \, \Phi^{\dag}_{N,k_y}({\bf{r}}') \:.
\end{equation}

The weak localization correction to conductivity has two contributions corresponding to 
the standard diagrams,\cite{Zuzin,Kachorovskii}
\begin{equation}\label{sigma}
\sigma = \sigma_a + \sigma_b \:. 
\end{equation}
The term $\sigma_a$ is given by 
\begin{equation}\label{sigmaI_r}
\sigma_a = \frac{2 \hbar}{\pi} \int  \operatorname{Tr} \left[ F ({\bf{r}},{\bf{r}}') \, \mathcal{C}^{(3)}({\bf{r}}',{\bf{r}})  \right]
   d{\bf{r}}d{\bf{r}}' \:,
\end{equation}
where $F({\bf{r}},{\bf{r}}') = J_x({\bf{r}},{\bf{r}}') \otimes J_x({\bf{r}},{\bf{r}}')$, ${\bf J} ({\bf{r}},{\bf{r}}')$ 
is the current vertex, $\mathcal{C}^{(3)}({\bf{r}}',{\bf{r}}) = \int P({\bf r},{\bf r}_1) \, \mathcal{C}({\bf{r}}_1,{\bf{r}}') d {\bf r}_1$, 
and the valley and spin degeneracy is taken into account. In graphene, the vertex can be 
presented in the form
\begin{equation}\label{vertex}
{\bf J} ({\bf{r}},{\bf{r}}') = 2 e v \int G^R({\bf{r}},{\bf{r}}_1) \,\boldsymbol{\sigma}\, G^A({\bf{r}}_1,{\bf{r}}') d {\bf r}_1  \:,
\end{equation}
where the factor 2 stems from the difference between the quantum and transport 
relaxation times. Straightforward calculation shows that the conductivity 
correction $\sigma_a$ assumes the form
\begin{equation}\label{sigmaI}
\sigma_a = - \frac{8 e^2}{\pi^2\hbar} \frac{\ell^2}{\ell_B^2} \mathrm{Tr} \sum_{N = 0}^{\infty}  \left[ \Pi \left(\mathbb{I}-P_N\right)^{-1} P_N^3 \right] \:,
\end{equation}
where 
\[
 \Pi = 
   \left(\begin{array}{cccc} 
   1 &  0 & 0 & 0 \\
   0 & -1 & 0 & 0 \\
   0 &  0 & 1 & 0 \\
   0 &  0 & 0 & 1  \\
  \end{array}\right) \:.
\]

Similar procedure can be carried out to derive the conductivity correction $\sigma_b$ 
corresponding to nonbackscattering interference effects. The calculation is more 
complicated, because it requires the expansion of each of the current vertices 
$\bf{ J}({\bf r}, {\bf r}')$ in the series of $\Phi_{N,k_y}$, and yields
\begin{equation}\label{sigmaII} \begin{split}
\sigma_b &=
  \frac{4 e^2}{\pi^2\hbar} \frac{\ell^2}{\ell_B^2}   
    \operatorname{Tr} \Bigg\{
      Q_0^T \Pi Q_0 (\mathbb{I}-P_0)^{-1}P_0 
\\
   & + \sum_{N=0}^{\infty}\left[
        Q_{N} \Pi Q_{N}^T + Q_{N+1}^T \Pi Q_{N+1} 
      \right] (\mathbb{I}-P_N)^{-1} P_N 
\Bigg\},
\end{split}
\end{equation}
where $Q_N$ are the matrices (here we assume $P_N^{(M)}=0$ for $M>N$)
\begin{subequations}\label{Qn}
\begin{equation}
Q_{N \geq 1 }= \left( \begin{array}{cccc}
   0 & 0 & 0 & 0 \\
   0 & P_N^{(1)}                     & -i\dfrac{P_N^{(2)}}{\sqrt2}  & -i\dfrac{P_N^{(0)}}{\sqrt2}  \\
   0 & i\dfrac{P_{N-1}^{(0)}}{\sqrt2} & \dfrac{P_{N-1}^{(1)}}2       & -\dfrac{P_{N}^{(1)}}2 \\
   0 & i\dfrac{P_{N+1}^{(2)}}{\sqrt2} & \dfrac{P_{N+1}^{(3)}}2       &   \dfrac{P_{N+1}^{(1)}}2\\
\end{array}\right) ,
\end{equation}
\begin{equation}
Q_0= \left( \begin{array}{cccc}
  0 & 0 & 0 & 0 \\
  0 & 0 & -i\dfrac{P_0^{(0)}}{\sqrt2}  & 0  \\
  0 & 0 & 0                       & 0 \\
  0 & 0 & \dfrac{P_1^{(1)}}2           & 0  \\
\end{array}\right) .
\end{equation}
\end{subequations}

Equations~\eqref{sigma}, \eqref{sigmaI}, and~\eqref{sigmaII} describe the quantum 
corrections to conductivity in the whole range of classically weak magnetic fields.

\section{Results and discussion}
Figure~1 presents the magnetic field dependence of the conductivity correction 
$\sigma(B)$ calculated for different ratios between the relaxation time $\tau$ 
and phase breaking time $\tau_{\phi}$. One can see that the conductivity correction 
is positive in the whole range of magnetic fields (weak antilocalization) and 
monotonously decreases with $B$ giving rise to negative magnetoconductivity. Such a 
behavior is in accordance with the qualitative analysis of intravalley interference 
effects presented above. The magnetic field dependences of the contributions $\sigma_a$ 
and $\sigma_b$ are plotted in Fig.~2 illustrating that the contributions are of opposite 
sign and comparable in magnitude.

\begin{figure}[t]
\includegraphics[width=\linewidth]{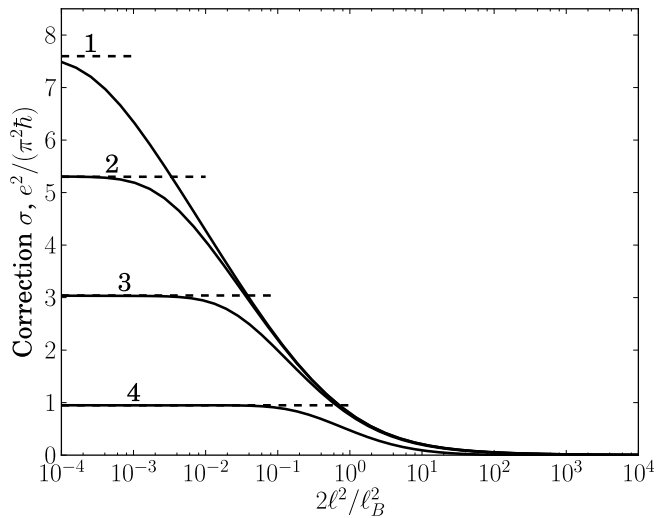}
\caption{Weak localization correction to conductivity as a function of magnetic 
  field. Curves 1, 2, 3, and 4 are plotted for $\tau/\tau_{\phi}=10^{-4}$, $10^{-3}$, 
  $10^{-2}$, and $10^{-1}$, respectively. Dashed lines show the weak-localization 
  corrections to conductivity at zero magnetic field
}
\end{figure}

Equations~\eqref{sigmaI} and~\eqref{sigmaII} allow us to analyze the behavior of 
conductivity in low and high magnetic fields. In the low-field limit, the contributions 
to magnetoconductivity assume the form   
\begin{eqnarray}\label{diffusion}
&& \sigma_a(B) - \sigma_a(0)
= - \frac{2 e^2}{\pi^2\hbar} F_2 \left( \frac{4 \tau_{\phi}}{\tau} \frac{\ell^2}{\ell_B^2} \right) \:, \\
&& \sigma_b(B)-\sigma_b(0)
= \frac{e^2}{\pi^2\hbar} F_2 \left( \frac{4 \tau_{\phi}}{\tau} \frac{\ell^2}{\ell_B^2} \right) \:, \nonumber
\end{eqnarray}
where $F_2(x)=\ln x + \psi(1/2+1/x)$, $\psi$ is the digamma function. These dependences 
are in agreement with the result of Ref.~\onlinecite{McCann} obtained in the diffusion 
approximation. In the inset of Fig.~2, we compare the magnetoconducitivity 
$\sigma(B) - \sigma(0)$ calculated after accurate Eqs.~(\ref{sigmaI}) and~(\ref{sigmaII})
(solid curves) with that calculated in the diffusion approximation Eqs.~(\ref{diffusion})
(dashes curves). One can see that the diffusion approximation describes the 
magnetoconductivity in low fields, where $\ell_B > \ell$. In higher magnetic fields, the 
diffusion approximation is not valid and one has to use the microscopic theory developed 
in this paper to describe the magnetoconductivity. Particularly, the ratio 2:1 between 
$|\sigma_a(B)-\sigma_a(0)|$ and $|\sigma_b(B)-\sigma_b(0)|$  obtained in the diffusion 
approximation does not hold any more. In the high-field limit, i.e., $\ell_B \ll \ell$, 
we may keep only first order in $\ell_B/\ell$ terms in Eq.~\eqref{PNM}, which gives 
$\sigma \propto 1/\sqrt{B}$. 
\begin{figure}[t]
\includegraphics[width=\linewidth]{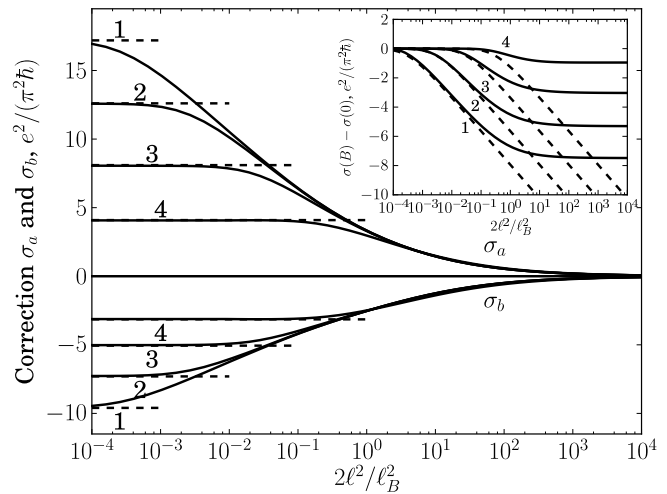}
\caption{
  Magnetic field dependences of the contributions $\sigma_a$ and $\sigma_b$. 
  Curves are plotted for the same ratios $\tau/\tau_{\phi}$ as in Fig.~1.
  Inset illustrates the dependences $\sigma(B)-\sigma(0)$ on magnetic field
  calculated after accurate Eqs.~(\ref{sigmaI}) and~(\ref{sigmaII}) (solid curves) and
  in the diffusion approximation (dashes curves)
}
\end{figure}

The absolute value of the weak-localization correction to conductivity in zero magnetic 
field can be obtained from Eqs.~\eqref{sigmaI} and~\eqref{sigmaII}  by taking a 
formal limit $\ell/\ell_B \rightarrow 0$. The calculated $\sigma(0)$, $\sigma_a(0)$, 
and $\sigma_b(0)$ are shown in Figs.~1 and~2 by dashed lines. The correction $\sigma(0)$ 
is determined by the phase breaking time, which can be used to experimentally study
the dependence of $\tau_{\phi}$  upon temperature or carrier density.    

Finally, we note that the presented theory can be applied to describe 
the weak localization of two-dimensional Dirac fermions in conducting surface of bulk 
topological insulators. In such systems, there is one two-dimensional channel and the 
conductivity corrections are given by Eqs.~\eqref{sigmaI} and~\eqref{sigmaII} divided by $4$.

To summarize, we have developed the theory of weak localization for graphene beyond the 
diffusion regime and calculated the interference corrections to conductivity in the whole
range of classically weak magnetic fields. The theory will allow one to better describe 
experimental data and more precisely determine the phase relaxation time of electrons in 
graphene.

\begin{acknowledgments}
  The authors acknowledge fruitful discussions with L.E.~Golub, M.M.~Glazov, and 
  M.I.~Dyakonov. 
  This work was supported by the RFBR, Russian President grant for young scientists, 
  and the Foundation ``Dynasty''-ICFPM.
  Partly supported by the Russian Ministry of Education and Science 
  Contract  (N14.740.11.0892)
\end{acknowledgments}


\end{document}